\documentclass[12pt]{article}
\usepackage{amssymb,slashed,latexsym,ifpdf,amsmath}

\usepackage[nosort]{cite}

\newcommand{\ft}[2]{{\textstyle\frac{#1}{#2}}}

\def\rme{{\rm e}}

\def\K{K\"{a}hler}
\newsavebox{\uuunit}
\sbox{\uuunit}
    {\setlength{\unitlength}{0.825em}
     \begin{picture}(0.6,0.7)
        \thinlines
        \put(0,0){\line(1,0){0.5}}
        \put(0.15,0){\line(0,1){0.7}}
        \put(0.35,0){\line(0,1){0.8}}
       \multiput(0.3,0.8)(-0.04,-0.02){12}{\rule{0.5pt}{0.5pt}}
     \end {picture}}

\newcommand{\bbox}{\lower.2ex\hbox{$\Box$}}
\csname @addtoreset\endcsname{equation}{section}

\newcommand{\rf}[1]{(\ref{#1})}
\newcommand{\vp}{\varphi}

\def\aD3{{\overline {\rm D3}}}
\def\be{\begin{equation}}
\def\ee{\end{equation}}
\def\ba{\begin{array}}
\def\ea{\end{array}}
\def\bea{\begin{eqnarray}}
\def\eea{\end{eqnarray}}

\def\K{K{\"a}hler}

\usepackage{color}
\definecolor{darkgreen}{rgb}{0,.5,0}

\newcommand{\BPhi}{\mathbf{\Phi}}
\newcommand{\BPhibar}{\mathbf{\overline \Phi}}
\newcommand{\BS}{\mathbf{S}}
\newcommand{\BSbar}{\mathbf{\overline S}}

\newcommand{\thetabar}{\bar \theta}
\newcommand{\chibar}{\overline \chi}
\newcommand{\psibar}{\bar \psi}
\newcommand{\sigmabar}{\overline \sigma}

\newcommand{\Dbar}{\overline D}
\newcommand{\nablabar}{\overline \nabla}

\newcommand{\Fbar}{\overline F}
\newcommand{\Wbar}{\overline W}

\newcommand{\ibar}{\bar i}
\newcommand{\jbar}{\bar j}
\newcommand{\ybar}{\bar y}
\newcommand{\zbar}{\bar z}

\newcommand{\Phibar}{\overline \Phi}
\newcommand{\sbar}{\bar s}
\newcommand{\Sbar}{\overline S}

\newcommand{\BA}{\mathbf{A}}
\newcommand{\BB}{\mathbf{B}}
\newcommand{\BLam}{\mathbf{\Lambda}}
\newcommand{\BOmega}{\mathbf{\Omega}}

\newcommand{\ga}{\bar{\gamma}}

\begin{document}

\begin{titlepage}

\begin{center}

\begin{flushright}
MIT-CTP 4741\\
CERN-PH-TH-2015-274
\end{flushright}

\vskip 2cm

{\Large {\bf  Cosmology with orthogonal nilpotent superfields}}

\vskip 1cm

 {\large Sergio Ferrara$^{1,2,3}$},   {\large Renata Kallosh$^4$},   {\large Jesse Thaler$^5$} \vskip 0.8cm
{\small\sl\noindent
$^1$ Theoretical Physics Department, CERN CHÐ1211 Geneva 23, Switzerland\\\smallskip
$^2$ INFN - Laboratori Nazionali di Frascati Via Enrico Fermi 40, I-00044 Frascati, Italy\\\smallskip
$^3$ Department of Physics and Astronomy, U.C.L.A., Los Angeles CA 90095-1547, USA\\\smallskip
$^4$ SITP and Department of Physics, Stanford University, Stanford, CA 94305, USA \\\smallskip
$^5$ Center for Theoretical Physics, Massachusetts Institute of Technology, Cambridge, MA 02139, USA}

\vskip 2cm
\begin{center}
{\bf Abstract}
\end{center}

\end{center}
{\small
We study the application of a supersymmetric model with two constrained supermultiplets to inflationary cosmology.  The first superfield $\BS$ is a stabilizer chiral superfield satisfying a nilpotency condition of degree 2, $\BS^2=0$.  The second superfield $\BPhi$ is the inflaton chiral superfield, which can be combined into a real superfield $\BB\equiv {1\over 2 i}\Big ( \BPhi-\BPhibar\Big)$.  The real superfield $\BB$ is orthogonal to $\BS$, $\BS \BB=0$, and satisfies a nilpotency condition of degree 3, $\BB^3=0$.  We show that these constraints remove from the  spectrum the complex scalar sgoldstino, the real scalar inflaton partner (i.e.\ the ``sinflaton''), and the fermionic inflatino.  The corresponding supergravity model with de Sitter vacua describes a graviton, a massive gravitino, and one real scalar  inflaton, with both the goldstino and inflatino being absent in unitary gauge.  We also discuss relaxed superfield constraints where $\BS^2=0$ and $\BS  \BPhibar$ is chiral, which removes the sgoldstino and inflatino, but leaves the sinflaton in the spectrum.  The cosmological model building in both of these inflatino-less models offers some advantages over existing constructions.
 }


\

\vspace{2mm} \vfill \hrule width 3.cm
{\footnotesize \noindent e-mails: Sergio.Ferrara@cern.ch, kallosh@stanford.edu, jthaler@mit.edu }
\end{titlepage}
\addtocounter{page}{1}
 \tableofcontents{}
\newpage

\section{Introduction}

De Sitter space plays a crucial role in our current understanding of cosmological observations \cite{Ade:2015xua}.  The inflationary epoch in the early universe occurs in approximate de Sitter space, and the present day acceleration of the universe will lead to de Sitter space asymptotically.  If supersymmetry is realized in nature, then it must be spontaneously broken during any de Sitter phase.  It is therefore interesting to study de Sitter vacua with spontaneously broken local supersymmetry, especially in cases where the interaction of matter with gravity is an essential ingredient.

In a cosmological context, one can ask whether supersymmetry is linearly or nonlinearly realized in a de Sitter phase.  Standard linear multiplets involve partner particles which differ by a half-unit of spin \cite{Golfand:1971iw}:  spin-0 scalars with spin-1/2 fermions, spin-1/2 fermions with spin-1 vectors, spin-3/2 gravitino with spin-2 graviton, and so on.  These supermultiplets form representations of the superalgebra and the supersymmetry operation flips the spin of the state by 1/2.  It was first recognized in \cite{Volkov:1972jx} by Volkov-Akulov (VA) that spontaneously broken global supersymmetry can also be realized nonlinearly.  In such cases, the corresponding goldstino multiplet contains a spin-1/2 fermion, but it does not contain a spin-0 or spin-1 partner; rather, a nonlinear supersymmetry operation flips a 1-particle fermion state into a 2-particle fermion state.  It was recognized a long time ago in \cite{rocek} that nonlinear realizations of supersymmetry can also be described using constrained superfields, and interest in constrained multiplets was renewed in \cite{Komargodski:2009rz,Kuzenko:2010ef}.  While constrained multiplets were originally introduced for global supersymmetry, they can be consistently generalized to local supersymmetry, as will be explained below.  

In this paper, we show how single-field inflation can be consistently embedded in local supergravity with the help of constrained multiplets.  The resulting supergravity action is surprisingly economical, since it describes the dynamics of just a single real inflaton, a graviton, and a massive gravitino.  Our construction uses the superfield content of \cite{Kahn:2015mla}, with two constrained chiral multiplets $\BS$ and $\BPhi$.  The stabilizer field $\BS$ satisfies a nilpotency condition of degree 2,\footnote{We use the terms ``nilpotent'' and degree of nilpotency as if superfields were square matrices:  a superfield $\mathbf{X} (x, \theta, \thetabar)$ is nilpotent of degree $r$ if $r$ is the least positive integer such that $\mathbf{X}^r=0$.  The term ``orthogonal'' will be used here as in the case of vectors:  two superfields $\mathbf{X}$ and $\mathbf{Y}$ are orthogonal if $\mathbf{X} \mathbf{Y} =0$.}
\be
\label{eq:nilpotentintro}
\BS^2 = 0,
\ee
and the inflaton is embedded in an orthogonal multiplet $\BPhi$ satisfying
\be
\label{eq:orthogonalityintro}
\BS \BB = 0, \qquad \BB\equiv \frac{1}{2 i} \Big ( \BPhi-\BPhibar\Big).
\ee
The real superfield $\BB$ also satisfies a nilpotency condition of degree 3, $\BB^3=0$ \cite{Komargodski:2009rz}.  Naively, a chiral inflaton muliplet $\BPhi$ would contain a fermionic inflatino partner and an additional real ``sinflaton'' scalar partner,\footnote{We use the axion-saxion naming convention here.  Together, the real inflation and real sinflation form a complex scalar which appears in the lowest component of $\BPhi$.} but such states are absent in the nonlinear realization.  As we will explain, this presents new opportunities for inflationary model building, since cosmological challenges presented by the inflatino and sinflaton are now absent.  

The importance of the $\BS^2 = 0$ constraint for cosmology is already well known.  The unconstrained linear $\BS$ multiplet has components
\be
\BS = (S, \chi^s,F^s),
\ee
where $\chi^s$ is the goldstino, $S$ is its scalar partner (i.e.\ the sgoldstino), and the $F^s$ auxiliary field is the order parameter for supersymmetry breaking.  While there have been attempts to identify the sgoldstino with the inflaton \cite{AlvarezGaume:2010rt}, sgoldstino inflation scenarios face a number of challenges \cite{Achucarro:2012hg}.  Therefore, it is typically necessary to stabilize $S$ and decouple it from cosmological evolution.  For example, in the case of a supergravity version of the Starobinsky model, one can use a linear realization of supersymmetry \cite{Cecotti:1987sa} and stabilize the sgoldstino via a \K\ potential term $- c (\BS \BSbar)^2$ \cite{Kallosh:2013xya}.  More economically and elegantly, though, one can use a nonlinear realization of supersymmetry and simply remove the sgoldstino via the $\BS^2=0$ constraint \cite{Antoniadis:2014oya} and in this way one can build the bosonic part of Volkov-Akulov-Starobinsky supergravity.\footnote{In \cite{Antoniadis:2014oya}, supersymmetry is restored at the minimum of the potential, though, and the constrained goldstino multiplet $\BS$ becomes singular.}  

A proposal to use the nilpotent multiplet in a general class of cosmological models was made in \cite{Ferrara:2014kva}, starting with Lagrange multipliers in the superconformal models underlying supergravity.  The stabilizer field $\BS$ with $\BS^2 = 0$ is now known as a nilpotent multiplet, and the resulting inflationary models are known as ``sgoldstino-less'' models.  By now, there are various cosmological inflationary models consistent with the data and yielding supersymmetry breaking at the minimum of the potential \cite{Kallosh:2014via,Dall'Agata:2014oka,Carrasco:2015uma,Carrasco:2015rva}.    Many recent developments on sgoldstino-less models were described in \cite{Ferrara:2015cwa,Dall'Agata:2015zla}, as well as other directions of work with constrained superfields.
There are also supersymmetric effective field theories (EFTs) of inflation \cite{Cheung:2007st,Senatore:2010wk} built using nilpotent multiplets \cite{Baumann:2011nk,Kahn:2015mla}.

It is important to mention that the $\BS^2=0$ nilpotency condition is of more general interest beyond inflationary cosmology.  Unlike the gravitino mass parameter $m_{3/2}$ which contributes universally to a negative cosmological constant in local supersymmetry, the $\BS$ multiplet always contributes a positive cosmological constant, so it naturally appears in studies of de Sitter space.  The minimal case of local supersymmetry coupled just to $\BS$ leads to a nonlinear realization of pure de Sitter supergravity without low energy scalars \cite{Bergshoeff:2015tra,Hasegawa:2015bza}. The action is complete with all higher order fermion couplings, and this complete locally supersymmetric action can be subsequently coupled to matter fields \cite{Hasegawa:2015bza,Kallosh:2015tea,Kallosh:2015pho}.  In string theory, nilpotent multiplets arise \cite{Bergshoeff:2015jxa,Kallosh:2015nia} when constructing supersymmetric versions of KKLT uplifting to de Sitter vacua \cite{Kachru:2003aw}.\footnote{The idea that open string theory and spontaneous symmetry breaking may be associated with branes and non-linearly realized supersymmetry were discussed a while ago \cite{Polchinski:1998rq,Sugimoto:1999tx,Dudas:2000nv}, but only recently were explicit constructions  presented.}  The nilpotent multiplet plays an important role in KKLT and LVS moduli stabilization scenarios, in particular for applications to particle phenomenology and cosmology \cite{Aparicio:2015psl}.  Nilpotent multiplets also appear in studies of supersymmetry breaking with multiple ``goldstini'' \cite{Cheung:2010mc}.  In our view, the fact that the complete VA nonlinear goldstino action appears on the $D$-brane world-volume in string theory \cite{Bergshoeff:2015jxa} suggests that the consideration of constrained versus unconstrained multiplets (and linear versus nonlinear realizations) is broader issue than originally envisaged.

The goal of this paper is apply the logic of nonlinear realizations and superfield constraints to the inflaton itself.   The unconstrained inflaton multiplet $\BPhi$ has components
\be
\BPhi = (\vp + i b, \chi^\phi, F^\phi),
\ee
where $\vp$ is the inflaton, $b$ is the sinflaton, $\chi^\phi$ is the inflatino, and $F^\phi$ is an auxiliary field.  Just as the nilpotency constraint $\BS^2 = 0$ projects out the sgoldstino $S$ and replaces it with a goldstino bilinear term, the orthogonality constraint $\BS \BB = 0$ implies that $b$, $\chi^\phi$, and $F^\phi$ are no longer independent fields and are instead functionals of $\chi^s$, $F^s$, and $\vp$.  The fact that superfield constraints reduce the total number of physical low energy degrees of freedom was one of the motivations for the study of such constrained systems.  The relevance of the constrained $\BS$ and $\BPhi$ fields for cosmological applications was described in \cite{Kahn:2015mla} in the context of building a minimal supersymmetric EFT for fluctuations about a fixed inflationary background.  Here, we are interested in studying the full inflationary dynamics, including the end of inflation.

For cosmological applications, the key feature of these orthogonal nilpotent superfields is that they have a particularly simple form when $\chi^s=0$.  In a locally supersymmetric action, one can make a choice of unitary gauge for the gravitino where $\chi^s=0$.  We will show that in this gauge
\be
\chi^s=0\,  \qquad \Rightarrow \qquad S=b=\chi^\phi= F^\phi=0,
\ee
such that the sgoldstino, sinflaton, inflatino, and inflaton auxiliary field are not just constrained, but entirely absent from the action.  This presents an extraordinary opportunity for cosmology.   First, in cosmological models with 2 unconstrained superfields, one has to work hard to stabilize three of the scalars---$\text{Re}(S)$, $\text{Im}(S)$, and $b$---since cosmological data favors a single scalar field inflaton $\varphi$.  The constraints $\BS^2 = 0$, $\BS \BB=0$ automatically project out these unwanted scalar modes while maintaining a nonlinear realization of supersymmetry.  Second, at the end of inflation in the presence of two chiral fermions $\chi^s$ and $\chi^\phi$, there is a problem of gravitino-inflatino mixing \cite{Kallosh:2000ve,Nilles:2001ry,Nilles:2001fg}, which makes the study of matter creation in the early universe very complicated.  Supergravity models based on orthogonal nilpotent superfields have no inflatino in unitary gauge and therefore no gravitino-inflatino mixing.  These are obvious advantages for cosmology. 

In addition, the fact that $F^\phi=0$ means the scalar potential of these models is different from ones studied in the past.  In the simplest case with a canonically-normalized inflaton, we will show that the scalar potential takes the form
\be
V(\phi) = f^2 (\vp) - 3 \kappa^2 g^2(\vp)= f^2 (\vp) - 3 M_{\rm Pl}^2 m^2_{3/2}(\vp),
\ee
where $\kappa = \sqrt{8 \pi G} = M_{\rm Pl}^{-1}$, $f(\varphi)$ is related to the degree of supersymmetry breaking, and $g(\varphi)$ gives rise to a field-dependent gravitino mass $m_{3/2}(\vp)$.  Surprisingly, $g'(\varphi)$ is absent from the potential, which appears to be a generic prediction of inflatino-less constructions using constrained multiplets.  As shown in \cite{Carrasco:2015iij}, this feature simplifies the construction of viable cosmological models.  Here, our focus is showing that inflatino-less models in de Sitter supergravity are internally consistent.

The remainder of this paper is organized as follows.  We review the structure of orthogonal nilpotent superfields in global supersymmetry in section \ref{sec:globalcase} and then generalize it to local supersymmetry in section \ref{sec:globaltolocal}.  We highlight a counterintuitive feature of the scalar potential in section \ref{sec:sugrapotential} and discuss generic aspects of inflatino-less models in \ref{sec:inflatinolessmodels}.  We provide alternative inflatino-less constructions in section \ref{sec:alternative} and conclude in section \ref{sec:discussion}.

\section{Orthogonal nilpotent superfields: $\mathbf{\BS^2=0}$, $\mathbf{\BS \BB=0}$}
\label{sec:globalcase}

\subsection{Structure in global superspace}

We start our discussion in global superspace.  Consider a chiral superfield $\Dbar_{\dot \alpha} \BS=0$ which is nilpotent of degree 2.
In the chiral basis $y^\mu= x^\mu + i \theta \sigma^\mu \thetabar$ where 
\be
\BS(y^\mu, \theta) = S(y) +  \sqrt 2 \theta \chi^s(y) + \theta^2 F^s(y),
\label{S} \ee
 the second degree nilpotency condition 
\be
\BS^2(y, \theta)=0
\ee
leads to 3 constraint equations involving the complex scalar sgoldstino $S(x)$, the fermionic field goldstino $\chi^s(x)$,\footnote{In local supersymmetry models with many matter multiplets, the name goldstino is typically reserved for the combination $v$ of various spin $1/2$ fields interacting with gravitino via $\psi_\mu \gamma^\mu v$, see section \ref{subsec:unitarygauge}.  The $\chi^s$ field will have this property, which justifies the name here.} and the auxiliary field $F^s(x)$:
 \be
S^2=0, \qquad  S\chi^s=0, \qquad S=  \frac{(\chi^s)^2}{2F^s}.
\label{S2}
\ee
Because the sgoldstino has been removed from the low energy spectrum, this nilpotent superfield has proved to be very useful in constructing viable cosmological models in the framework of 2-superfield models \cite{Kallosh:2014via,Dall'Agata:2014oka,Carrasco:2015uma,Carrasco:2015rva,Ferrara:2015cwa,Dall'Agata:2015zla}.

Following  \cite{Kahn:2015mla}, we introduce a second chiral multiplet $\BPhi$ 
\be
\BPhi(y^\mu, \theta) = \Phi(y)+  \sqrt 2 \theta \chi^\phi(y) + \theta^2 F^\phi(y)
\label{Phi} \ee 
with 
\be
\Phi(x)=  \vp(x)+i b(x).
\ee 
Here, the real scalar $\vp(x)$ is the inflaton, the real scalar $b(x)$ is its sinflaton partner, the fermion $\chi^\phi $ is the inflatino, and $F^\phi$ is the auxiliary field.  From the anti-chiral superfield
 \be
\BPhibar(\ybar^\mu, \theta) =\Phibar (\ybar)  +  \sqrt 2 \thetabar \chibar^\phi(\ybar) + \thetabar^2 \Fbar^{\phi}(\ybar),
 \ee 
we can construct the real superfield $\BB$,
\be
\BB(x, \theta, \thetabar)\equiv \frac{1}{2 i}\Big ( \BPhi-\BPhibar\Big).
\ee
This $\BB$ field can be defined in a basis where the chiral superfield is short, and the anti-chiral superfield is long.  Namely, we keep \rf{S} and \rf{Phi} but rewrite $\BPhibar$ as
 \begin{align}
\BPhibar( y^\mu- 2i \theta \sigma^\mu \thetabar, \thetabar) &= \Phibar(y)  +  \sqrt 2 \thetabar \chibar^\phi( y) + \thetabar^2 \Fbar^{\phi}( y) + \ldots
 \end{align}
 We impose an orthogonality constraint on $\BB$ via
 \be
 \label{eq:orthogonality}
\BS(y, \theta)  \BB(y, \theta, \thetabar) =0.
 \ee
 
This orthogonality relation produces a number of constraint equations for the component fields, for each of the  $\theta^m \thetabar^n$ with $0\leq m\leq 2$ and $0\leq n\leq 2$. By solving these equations one finds the following \cite{Komargodski:2009rz,Kahn:2015mla}:\footnote{Because \cite{Kahn:2015mla} only considered fluctuations about a fixed inflationary background, derivative terms on $F$ were neglected.}  
 \begin{enumerate}
\item The first component of the inflaton multiplet has a real scalar field inflaton $\vp$ as well as fermionic $\chi^s$-dependent terms:
\begin{align}
\label{scalar}
\Phi & =  \varphi + \frac{i}{2}{\chi^s\over F^s} \sigma^\mu {\chibar^s\over \Fbar^s} \partial_\mu \varphi + \frac{1}{8} \left (\Big({\chi^s\over F^s}\Big)^2 \partial_\nu \Big({\chibar^s\over \Fbar^s} \Big) \sigmabar^\mu \sigma^\nu {\chibar^s\over \Fbar^s}  - \text{c.c.} \right)\partial_\mu \varphi \\ \nonumber
& \qquad ~ -\frac{i}{32}\Big ({\chi^s\over F^s}\Big)^2 \Big({\chibar^s\over \Fbar^s} \Big)^2 \partial_\mu  \Big ({\chibar^s\over \Fbar^s} \Big)  (\sigmabar^\rho \sigma^\mu \sigmabar^\nu + \sigmabar^\mu \sigma^\nu \sigmabar^\rho) \partial_\nu \Big({\chi^s\over F^s} \Big) \partial_\rho \varphi.
\end{align}
Note that there is no independent scalar sinflaton $b(x)$ in this expression, and it has been replaced by terms bilinear or higher in the fermion $\chi^s$.  This is desirable from the perspective of cosmology, since there is no need to worry about this sinflaton mode being too light (or tachyonic).  This is the analogous feature that we saw with the $\BS^2 = 0$ nilpotency condition; the sgoldstino $S$ no longer contributes to the bosonic evolution since it is constrained to be a function of the fermions $S= (\chi^s)^2 / (2F^s)$.  Moreover, when we transition to local supersymmetry in section \ref{sec:globaltolocal}, we can work in a unitary gauge for the gravitino where $\chi^s = 0$, in which case we will be able to show that $\Phi = \varphi$ is a pure real function.
\item The inflatino $\chi^\phi$ is no longer independent and is rather proportional to the goldstino $ \chi^s $:
 \be
 \chi^\phi = i \sigma^\mu  { \chibar^s \over   \, \Fbar^s} \partial_ \mu \Phi.
 \label{inflatino}
 \ee 
This striking feature of the orthogonality/nilpotency constraint might lead to a solution of the long-standing cosmological problem of gravitino-inflatino mixing which will be explained in detail in sections \ref{sec:globaltolocal} and \ref{sec:inflatinolessmodels}.  The importance of this is that in unitary gauge, both the inflatino $\chi^\phi$ and the goldstino $\chi^s$ fermion vanish and the only relevant fermionic degree of freedom is the massive gravitino.
\item  The auxiliary field is at least quadratic or higher in $\chi^s$:
 \be
 F^\phi=  -\partial_\nu \Big({\chibar^s\over \Fbar^s} \Big)  \sigmabar^\mu \sigma^\nu \Big({\chibar^s\over \Fbar^s} \Big)  \partial_\mu \Phi + \frac{1}{2}\Big({\chibar^s\over \Fbar^s} \Big)^2 \partial^2 \Phi .
\label{aux} \ee
 The fact that $F^\phi$ vanishes in unitary gauge in a locally supersymmetric model leads to an interesting conclusion that the dependence of the off-shell bosonic potential on $F^\phi$ is eliminated into the fermionic sector, as discussed in section \ref{sec:sugrapotential}.
 \end{enumerate}
 By direct computation, one can show that the nilpotency condition \rf{S2} and orthogonality condition \rf{eq:orthogonality} imply a 3rd degree nilpotency condition \cite{Komargodski:2009rz}
 \be
 \BB^3(x, \theta, \thetabar)=0.
 \ee
This can be understood since the imaginary part of $\Phi$ contains one undifferentiated 2-component spinor $\chi^s$, and therefore the product of 3 of them vanishes.

\subsection{Constructing an action}

The most general supersymmetric action for the $\BS$ and $\BPhi$ superfields at the 2-derivative level is
\be
\mathcal{L} = \int d^4 \theta \, K(\BS,  \BSbar; \BPhi ,  \BPhibar ) + \Big (\int d^2 \theta \, W (\BS, \BPhi) + \text{h.c.} \Big ).
\ee
In a cosmological context, it is often useful to impose an approximate shift symmetry on the inflaton $\varphi$, which is only broken by the holomorphic superpotential.  This can be accomplished by requiring that the \K\, potential has a manifest inflaton shift symmetry and depends only on the real superfield $\BB$ via (see also \cite{Kahn:2015mla}):
\be
K(\BS,  \BSbar; \BPhi ,  \BPhibar )= \BS  \BSbar   + \BB^2.
\label{K}
\ee
Here we neglect terms linear in $\BS$ and in $\BB$, since they may be removed by a \K\, transform.

The superpotential is a holomorphic function of the inflaton superfield $\BPhi$, therefore it cannot depend on the shift symmetric $\BB$ (which contains the anti-holomorphic $\BPhibar$). Any dependence of the superpotential on $\BPhi$ introduces deviation from the shift symmetry of the \K\ potential given in \rf{K}.\footnote{In \cite{Kahn:2015mla}, the superpotential was taken to be independent of the inflaton, leading to a flat inflaton potential.  In that case, $\langle \dot{\vp} \rangle$ must be inserted by hand, as expected since that EFT only aims to describe fluctuations about a fixed inflating background.}  The most general superpotential we can write is
\be
W (\BS, \BPhi)= f(\BPhi) \,  \BS + g (\BPhi).
\label{W}
\ee

Going beyond the assumption of a strict shift symmetry in \rf{K}, the most general form of the \K\ potential consistent with nilpotency and orthogonality is
\be
\label{eq:extendedK}
K(\BS,  \BSbar; \BPhi ,  \BPhibar ) = h_0 (\BA) \, \BS \BSbar  + h_1(\BA) +  h_2(\BA) \, \BB  + h_3(\BA) \, \BB^2, \qquad \BA \equiv {1\over 2}\Big ( \BPhi+ \BPhibar\Big).
\ee
However, this expression can typically be simplified using field redefinitions and \K\ transformations; the details are given in appendix \ref{app:generalK}.  Assuming the $h_i(\BA)$ functions are not pathological, 
\rf{eq:extendedK} can be reduced to
\be
\label{eq:extendedKfinal}
K(\BS,  \BSbar; \BPhi ,  \BPhibar ) = \BS \BSbar + h(\BA) \, \BB^2 .
\ee
The function $h(\varphi)$ generates a nonminimal \K\ metric for $\BPhi$ and introduces an additional breaking of the shift symmetry.  For simplicity, we set $h(\varphi) = 1$ for our discussion in section \ref{sec:inflatinolessmodels}, though it might lead to interesting features for cosmological model building.

With regards to model building, it is sometimes more convenient to start with general models in the form
\be
K( \BS, \BSbar;  \BPhi ,  \BPhibar)=  h_0 (\BA) \,  \BS \BSbar  + k( \BPhi ,  \BPhibar),  \qquad W= f(\BPhi) \,  \BS + g (\BPhi),  \qquad \BS^2= \BS \BB = 0,
\label{KW1}
\ee 
where $k( \BPhi ,  \BPhibar)$ might have a particular symmetry before the constraints are imposed and the term $h_0 (\BA)$ is present in string-inspired models with warped geometry. For example, $\alpha$-attractor models \cite{Carrasco:2015uma,Carrasco:2015rva} have a hyperbolic \K\ geometry encoded in $k( \BPhi ,  \BPhibar)$, and one might therefore be interested in using these geometric variables directly instead of performing a \K\ transform to simplify the \K\ potential.  Examples with warped geometry are given in section 4 of \cite{Kallosh:2015nia}, where $\BPhi$ is a multiplet representing the volume of an extra dimension.  Of course, \rf{eq:extendedK} and \rf{KW1} are physically equivalent, but using variables with particular physics interpretations may be preferable as a starting point.  Indeed, using both variable choices might give extra insights into cosmological model building.

\section{From global to local supersymmetry}  
\label{sec:globaltolocal}

In ordinary cosmological models with local supersymmetry and (at least) two chiral multiplets, the inflaton is an unconstrained chiral superfield and the inflatino is present in the low energy spectrum.\footnote{There are inflationary models built from just one chiral multiplet \cite{Goncharov:1985yu,Ketov:2014qha,Linde:2014ela}, in which case the inflatino and the goldstino are identified. In such models, there is no inflatino in unitary gauge and therefore no inflatino-gravitino mixing.  However, in these models it is very difficult to achieve supersymmetry breaking at the end of inflation \cite{Linde:2014ela} and a second field typically has to be introduced.}  There is only one combination of the spin 1/2 particles which can be removed by a gauge fixing condition, and the inflatino generically remains mixed with gravitino.  For this reason, the analysis of gravitino-inflatino production in models with two or more chiral multiplets is rather involved \cite{Kallosh:2000ve,Nilles:2001ry,Nilles:2001fg} (see section \ref{sec:inflatinolessmodels} below).  As anticipated in \rf{inflatino}, we now show how the orthogonality condition $\BS \BB = 0$ plus the appropriate gravitino gauge choice can remove these inflatino complications.

Compared to the previous section, we are now transitioning from 2-component Weyl notation to 4-component Majorana notation for the fermions in order to make contact with the supergravity literature.

\subsection{Importance of unitary gauge}
\label{subsec:unitarygauge}

Local supersymmetry, as any gauge symmetry, requires a choice of the gauge-fixing condition. The super-Higgs effect, where the gravitino $\psi_\mu$ becomes  massive by eating a goldstino $v$, can be explained by the fact that the action with  local supersymmetry requires a gravitino-goldstino mixing term
\be
 \psibar^\mu \gamma_\mu \,  v + \text{h.c.},
\ee
where the goldstino $v$ is some combination of spin $1/2$ fermions.  In the case of cosmological models with two superfields, a nilpotent stabilizer $\BS$ and an \emph{unconstrained} inflaton multiplet $\BPhi$, the goldstino field is  
\be
v =   {1\over \sqrt 2}  \rme^{K/2} (\chi^s \nabla_s W +  \chi^\phi \nabla_{\phi }W),  
\label{v}\ee
where $\nabla_i$ is a \K-covariant derivative with respect to the $i$-th chiral multiplet.  

The possible unitary gauges for models with a nilpotent multiplet and an unconstrained inflaton multiplet were discussed in \cite{Kallosh:2015sea}.  It turns out that the standard unitary gauge 
\be
v=0
\label{un}\ee
is not convenient for calculational purposes.  The reason is that the uneaten linear combination of $\chi^s$ and $\chi^\phi$ has complicated couplings inherited from the high powers of $\chi^s$ interactions necessitated by the $\BS^2=0$ constraint.  An alternative unitary gauge is\footnote{Any gauge-fixing condition of local supersymmetry which depends only on chiral matter fermions and not on the gravitino is a unitary gauge, in the sense that there are no propagating ghost degrees of freedom.  This is as opposed to gauges like $\gamma^\mu \psi_\mu=0$ or $D^\mu \psi_\mu=0$ which do necessitate propagating ghosts.}
\be
\chi^s=0,
\ee
which has some calculational advantages since it removes these higher order terms in $\chi^s$. The gauge, however, leaves a mixing term between the gravitino and inflatino,
\be
 \psibar^\mu \gamma_\mu    {1\over \sqrt 2}  \rme^{K/2}  \chi^\phi \nabla_{\phi }W.
 \label{GI}
 \ee
Note that one can set either $v=0$ or $\chi^s=0$ in these constructions, and one cannot realize both simultaneously unless $\nabla_{\phi}W = 0$.
 
In this context, imposing the orthogonality constraint $\BS \BB=0$ is a particularly attractive option, since it suggests the possibility of consistent ``inflatino-less'' cosmological models.  Recall from \rf{inflatino} that in global supersymmetry, the vanishing of $\chi^s$ implies the vanishing of inflatino:
\be
\chi^s=0 \qquad \Rightarrow \qquad \chi^\phi=0.
\label{prop}
\ee
If this relation were also to hold in local supersymmetry, then we could find a unitary gauge choice where 
\be
v=0\, \qquad {\rm and} \qquad \chi^s=0,
\ee 
despite the fact that local gauge symmetry naively allows only one of these conditions.  As we show below, \rf{prop} is indeed valid in local models, allowing us to achieve a very interesting class of inflatino-less cosmological models.

\subsection{Imposing the constraints}

In order to verify \rf{prop}, we need to figure out how the constraint $\BS \BB=0$ and the corresponding   equations \rf{scalar}, \rf{inflatino}, and \rf{aux} are modified in models with local supersymmetry.

In case of a single nilpotent constraint $\BS^2 = 0$, the generalization from global to local supersymmetry is known.  At the level of the superconformal theory, the constraint can be implemented via a Lagrange multiplier, a chiral superfield $\BLam$ \cite{Ferrara:2014kva}.  By supplementing the superconformal action with
\be
[\BLam \, \BS^2]_F + \text{h.c.},
\label{LM}
\ee
one can write down the superconformal action \rf{LM} in components, where the nilpotent chiral multiplet  is $(S, \chi^s, F^s)$ and the 
Lagrange multiplier is $(\Lambda, \chi^\Lambda, F^\Lambda)$. The corresponding locally superconformal action is given in \cite{Bergshoeff:2015tra}. The field equation for $\Lambda(x)$ in the local superconformal theory follows and is given by
\begin{equation}
   2S F^s - (\chi^s)^2 +\sqrt{2}\psibar_\mu \gamma^\mu S P_L\chi^s +\ft12 \psibar _{\mu }P_R \gamma ^{\mu \nu }\psi _{\nu }S ^2 =0\,.
 \label{fieldeqnLambdapre}
\end{equation}
By inspecting this equation together with similar equations for $\chi^\Lambda$ and $ F^\Lambda$, one finds that the solution of $\BS^2=0$ from \rf{S2} remains valid in the local theory, the most important one being
\begin{equation}
   2S F^s - (\chi^s)^2  =0\,.
 \label{fieldeqnLambda}
\end{equation}
We also see here that a nontrivial solution is possible only for $F^s\neq 0$.  This, in turn, means that the positive contribution to the cosmological constant $|F^s|^2$ in a model with local supersymmetry is a necessary consequence of the existence of a nontrivial nilpotent multiplet in a local theory.

The new orthogonality constraint $\BS\BB=0$ is no longer holomorphic.  This makes it more complicated to transition from the global theory to the local one, since it requires a $D$-term analysis in the superconformal theory.  First, it is straightforward to prove that by supplementing the superconformal action with a complex vector Lagrange multiplier $\BOmega$, 
\be
[\BOmega \, \BS \BB]_D + \text{h.c.},
\ee
the equations of motion for $\BOmega$ impose the superfield constraint $\BS \BB = 0$.\footnote{Alternatively, one can impose two real constraints, $(\BS + \BSbar) \BB = 0$ and $i (\BS - \BSbar) \BB = 0$, using two real vector Lagrange multiplier superfields.}  That said, it is a bit more involved to show how the constraints arising from $\BS \BB=0$ are modified in the local case, since this constraint equation now involves terms containing the gravitino and the vector auxiliary field of supergravity.

Before proceeding on this, it is worth reflecting on why the nilpotent constraint $\BS^2 = 0$ gives expressions \rf{S2} and \rf{fieldeqnLambda} which are equally valid in global and local supersymmetry.  The local superconformal calculus and the rules for multiplication of superfields are well known; for example, in the case of chiral multiplets, the comparison between the global and local rules is described in chapter 16.2.1 of \cite{Freedman:2012zz}.  In general, any ordinary derivative that appears in the global case must be replaced by a superconformal derivative in the local case (defined, e.g., in (16.34) and (16.37) of \cite{Freedman:2012zz}).  These superconformal derivatives introduce dependence on the gravitino and on the vector auxiliary field which were absent in the global case, in principle modifying the meaning of $\BS^2 = 0$.  Note, however, that all equations in \rf{S2} depend only on the undifferentiated spinor $\chi^s$ and undifferentiated scalar $S$.  Therefore, there is no place in these algebraic equations to replace ordinary derivatives with superconformal ones.  This is a shortcut to reach the conclusion that the global constraints in \rf{S2} are valid in the local theory.  Note that this same logic holds for any holomorphic constraint.\footnote{An interesting example of a holomorphic constraint is the chiral orthogonality condition $\BS \BPhi = 0$ \cite{Brignole:1997pe,Komargodski:2009rz}.  This removes the scalar modes from $\BPhi$ but leaves the auxiliary field $F^\phi$, as is relevant for describing more general matter interactions \cite{Dall'Agata:2015zla}.}

In the case of $\BS \BB=0$, on the other hand, the constraints depend on supercovariant derivatives.  This is expected since the global equations \rf{scalar}-\rf{aux} depend on ordinary derivatives.   One can see this explicitly in appendix \ref{app:orthogonalitylocal} where we derive the local version of $\BS \BB=0$ using the tensor calculus in supergravity \cite{Stelle:1978ye,Cremmer:1982wb}.  The supercovariant derivatives of $S$ and $\chi^s$ are
\begin{align}
\hat D_\mu S &= \partial_\mu S - \frac{i}{2} \psibar_\mu \chi^s_L, \\
 \hat{D}_\mu \chi^s_L & = D_\mu \chi^s_L - (\slashed{D} S) \psi_{\mu R} - F^s \psi_{\mu L} - \tfrac{i}{2} A_\mu \chi^s_L,
\end{align}
where $D_\mu \psi$ is an ordinary covariant derivative of a spinor including the spin connection, and $A_\mu$ is the vector auxiliary field.  These additional terms present in the local case significantly complicate the derivation of the constrained $\BPhi$ components.

\subsection{Simplification in unitary gauge}
\label{subsec:simplificationunitary}

Crucially, however, there is a simplification to the $\BS \BB=0$ constraint equation if we work in unitary gauge where $\chi^s = S = 0$.  In that case
\begin{align}
\label{eq:covderivunitarygauge}
\chi^s=0 \qquad \Rightarrow \qquad \hat D_\mu S &= 0, \\
\hat D_\mu \chi^s_L &= - F^s \psi_\mu.
\end{align}
At first glance, this might not seem like enough of a simplification, since $\hat D_\mu \chi^s_L$ still involves the gravitino and the (nonzero) auxiliary field $F^s$.  The key point, though, is that each fermionic term on the right-hand sides of \rf{scalar}-\rf{aux} contains as a factor at least one undifferentiated spinor $\chi^s$, which goes to zero in unitary gauge.  As shown explicitly in appendix \ref{app:orthogonalitylocal}, this undifferentiated $\chi^s$ remains as a factor in the local expressions as well.  So even if $\hat D_\mu \chi^s_L $ is nonvanishing, any terms that depend on it do vanish in this unitary gauge.

We therefore conclude that when the local supersymmetry is gauge fixed with $\chi^s=0$, the constrained values of the inflatino $\chi^\phi$ and auxiliary field $F^\phi$ vanish:
\be
\chi^s=0 \qquad \Rightarrow \qquad \chi^\phi = F^\phi=0.
\label{!}
\ee
Moreover, the scalar component of $\BPhi$ reduces to just a real inflaton:
\be
\Phi = \varphi.
\label{!!}
\ee
The above logic is validated by the explicit computation in appendix \ref{app:orthogonalitylocal}, which shows that nonzero terms involving the gravitino, which are indeed possible in principle, are in fact absent from the $\BS \BB=0$ constraint equations in unitary gauge.  This now opens the possibility of consistent inflatino-less cosmological models with local supersymmetry.

An interesting property of unitary gauge with vanishing inflatino and sinflaton is that the bosonic part of the vector auxiliary field 
\be
A_{\mu}^{\rm bosonic}={i\over 2} (K_{ i}  \partial_\mu  z^{ i}- K_{\ibar}  \partial_\mu \zbar^{\ibar})
\label{Amu}
\ee
vanishes for \K\ potentials of the form \rf{eq:extendedKfinal}, though not in the general case \rf{KW1} prior to performing a \K\ transform.

\section{Supergravity potential with constrained superfields}
\label{sec:sugrapotential}

In addition to the fact that the inflatino vanishes in unitary gauge, an interesting feature of \rf{!} is that the auxiliary field $F^\phi$ vanishes in this gauge as well.  In more general gauges, $F^\phi$ is nonzero but $\BS \BB=0$ still implies that $F^\phi$ does not have a bosonic part.  This leads to an interesting property of the bosonic action in these inflatino-less models.

\subsection{A nonstandard potential}

In ordinary supergravity with unconstrained multiplets, one would expect the supergravity potential with two chiral multiplets $\BS$ and $\BPhi$ to take the form 
\be
V= \rme^K (\nabla_i Wg^{i \ibar} \, \nablabar_{\ibar} \Wbar - 3 |W|^2), \qquad  i=s,\phi.
\label{st}
\ee
If the $\BS$ multiplet is nilpotent, the potential is still given by \rf{st}, simply evaluated at $S=0$.  The reason this works is that $F^s$ is a full auxiliary field whose value is determined by its equation of motion, just as in the unconstrained case.

By contrast, if the $\BPhi$ multiplet is orthogonal with $F^\phi$ being fermionic, the potential takes the unusual form
\be
\label{eq:strangeform}
V= \rme^K (\nabla_s Wg^{s\sbar} \, \nablabar_{\sbar} \Wbar - 3 |W|^2), \qquad \nabla_\phi W \neq 0.
\ee
That is, the term $\nabla_\phi W$ is not present in the potential, despite the fact that the bosonic part of $\nabla_\phi W$ is nonvanishing.  The intuitive reason for this result is that $F^\phi$ does not appear in the component action and has no equation of motion to contribute a $\nabla_\phi W$ term to the potential.

\subsection{Derivation from the off-shell action}

Because \rf{eq:strangeform} is sufficiently counter-intuitive, it deserves further explanation.  One way to explain it is to start with off-shell supergravity in the form given in \cite{Kallosh:2015tea},\footnote{We replace the index $\alpha$ labeling chiral multiplets in \cite{Kallosh:2015tea, Freedman:2012zz} by $i$ here.} namely
\be
\label{eq:Loffshell}
e^{-1} { \cal L}_{\rm off-shell}= (F^i - F^{i } _G  ) g_{i \ibar }    (\Fbar^{\ibar} - \Fbar^{\ibar } _G   ) + e^{-1} { \cal L}^{\rm book}\,,
\ee
where
\be\label{eq:F}
F^{i } _G \equiv  - \rme^{K/2} g^{i \ibar} \, \nablabar _{\ibar }\Wbar + (F_G^{i})^f\,.
\ee
In these expressions, $(F_G^{i})^f $ depends on fermions and ${ \cal L}^{\rm book}$ is the full supergravity action in \cite{Freedman:2012zz} with auxiliary fields eliminated on their equations of motion.  For unconstrained multiplets, the first term in \rf{eq:Loffshell} vanishes on the $F^i$ equations of motion.

Let us start with the case where $\BS$ is nilpotent but $\BPhi$ is unconstrained.  If we are interested only in the bosonic action, we can use the expressions above together with the constraint in \rf{S2} and deduce that the $F^s$ and $F^\phi$ equations of motion are still
\be
F^s=  -\rme^{K/2} g^{s \ibar} \, \nablabar _{\ibar }\Wbar, \qquad F^\phi=  -\rme^{K/2} g^{\phi \ibar} \, \nablabar _{\ibar }\Wbar,
\ee
except everywhere the condition 
\be
S=0
\ee
must be inserted.  The bosonic potential in this case is simply 
\be
V= \rme^K (\nabla_i W g^{i\ibar} \, \nablabar_{\ibar} \Wbar - 3 |W|^2)\Big |_{S=0}\qquad  i=s,\phi.
\label{st1}
\ee

Now consider the situation with an orthogonality constraint on $\BPhi$.  It is convenient to separate the action into terms that depend on the auxiliary fields,
 \be
{\cal L} (F, \Fbar) =(F^i - F^{i } _G  ) g_{i \ibar}    (\Fbar^{\ibar} - \Fbar^{\ibar } _G   )- F^{i } _G   g_{i \ibar }     \Fbar^{\ibar } _G,
\label{F}
\ee
and terms that are independent of $F^i$, 
\bea\label{Loffshell1}
 F^{i } _G   g_{i \ibar }     \Fbar^{\ibar } _G + e^{-1} { \cal L}^{\rm book}\,.
\eea
Note that the contribution to the bosonic potential proportional to $|\nabla_i W|^2$ is absent in \rf{Loffshell1}.  With the orthogonality condition on $\BPhi$, the new and unusual situation is that the off-shell auxiliary field $F^\phi$ is constrained to be
\be
F^\phi_{\rm bosonic}=0,
\label{f}
\ee
as implied by the local version of \rf{aux}.  Therefore $F^\phi$ does not contribute in \rf{F},
\be
{\cal L} (F, \Fbar)\Big |_{\rm bosonic} \Rightarrow F^s  g_{s \sbar }    \Fbar^{\sbar}  + \rme^{K/2}  ( F^s \nabla_{ s }{W}+ \nablabar _{\sbar } \Wbar  \Fbar^{\sbar}   ), 
\ee
and the total bosonic potential comes from integrating out $F^s$ and from the terms in \rf{Loffshell1}, leading to
\be
V= \rme^K (|\nabla_s W|^2 - 3 |W|^2)\Big |_{S=0},  \qquad \nabla_\phi W \neq 0,
\label{newpot}\ee
as claimed above and taking into account that $\BS^2=0$.

The argument above can also be given using the off-shell supergravity equations in the construction of \cite{Cremmer:1982wb}.  In that language, the part of the action relevant for the bosonic potential is
\be
-{1\over 3} \hat U \hat U^* + (\hat U W + \hat U^* \Wbar) \rme^{K/2} + K_{i\jbar} \hat h^i \hat h^{\jbar} + ( \hat h^i D_i W + \hat h^{\ibar} D_{\ibar} \Wbar) \rme^{K/2}.
\label{Sergio}
\ee
In the standard unconstrained situation, integrating out the compensator field $\hat U$ and the auxiliary fields $\hat{h}^i$ leaves us with the supergravity potential in \rf{st}.  If in this off-shell action we first take into account that $\hat h^\phi$ is fermionic, though, then the contribution of the term $\nabla_\phi W$ will drop out from the bosonic part of the action, leading to \rf{newpot}.

\section{Inflatino-less models with orthogonal nilpotent multiplets}
\label{sec:inflatinolessmodels}

As an example of the model building possibilities presented by inflatino-less constructions, consider supergravity models with a shift-symmetric \K\ potential as in \rf{K} and a generic superpotential as in \rf{W}:  \be
K( \BS, \BSbar;  \BPhi ,  \BPhibar)=  \BS \BSbar  - {1\over 4}  (\BPhi-\BPhibar)^2,  \qquad W= f(\BPhi) \,  \BS + g (\BPhi),  \qquad \BS^2= \BS \BB = 0.
\label{KW}
\ee 
In these models, violation of the inflaton shift symmetry enters only via the superpotential.

\subsection{Action in unitary gauge}
In the unitary gauge with $\chi^s=0$, the  action corresponding to \rf{KW} is surprisingly simple:
\begin{align}
 e^{-1}{\cal L} & = \frac{1}{2\kappa^2} \left[ R(\omega (e )) -\psibar_\mu \gamma ^{\mu \nu \rho } D_\nu \psi _\rho
 +{\cal L}_{\rm SG,torsion} \right] \nonumber\\
&\quad ~ +\frac{g(\vp)}{2}  \psibar_{\mu  } \gamma ^{\mu \nu }\psi _{\nu  } - {1\over 2} (\partial \vp)^2+3 \kappa^2 g^2(\vp) - f^2 (\vp).
\label{actionUG}
\end{align}
The field-dependent gravitino mass is 
\be
m_{3/2}(\vp) =  g(\vp) \kappa^2=  g(\vp) M_{\rm Pl}^{-2},
\ee
and we have taken the simple case where $g(\vp)$ and $f(\vp)$ are both real functions.  Here, the gravitino kinetic term is noncanonical in order to scale out the $\kappa$ coupling, the spinor derivative $D_\mu$ is a vierbein-dependent spin connection, and ${\cal L}_{\rm SG,torsion}$ is the standard quartic in gravitino term in ${\cal N}=1$ supergravity.  There are numbers of ways to confirm this action.  One way is to start from the results in \cite{Hasegawa:2015bza,Kallosh:2015tea}, set $\chi^s=0$ and $b = 0$, and remove the $(g')^2$ term from the potential as advocated in section \ref{sec:sugrapotential}.  An alternative approach is to use the explicit action in \cite{Kallosh:2015pho} following from the general formula given in \cite{Kallosh:2015tea}.  In particular the term proportional to $\epsilon^{\mu\nu\rho\sigma} \psibar_\mu\gamma_\nu\psi_\nu A_\rho$ with the vector auxiliary field vanishes in models based on \rf{KW}, as explained in \rf{Amu}.

The action in \rf{actionUG} describes the graviton, the gravitino with a field-dependent mass term proportional to $g(\varphi)$, a canonically normalized inflaton $\vp$, and an inflaton potential:
\be
V(\phi) = f^2 (\vp) - 3 \kappa^2 g^2(\vp)= f^2 (\vp) - 3 M_{\rm Pl}^2 m^2_{3/2}(\vp).
\ee
This potential has a de Sitter vacuum under the condition that $V>0$ at $V' \equiv \partial V / \partial \varphi =0$, namely
\be
f^2- 3 \kappa^2 g^2 > 0 \qquad {\rm at} \qquad  ff'- 3 \kappa^2 gg'=0.
\ee
There is no sinflaton, sgoldstino, or inflatino in the spectrum.  

We can also look at general models in the form of \rf{KW1} where the geometry of the moduli space is important but with canonical $\BS \BSbar$, using \cite{Hasegawa:2015bza,Kallosh:2015tea,Kallosh:2015pho}.  In such cases, there will be a few corrections to the action in \rf{actionUG}:  the kinetic term of the real scalar will include dependence on $k_{\Phi \Phibar}$, the potential might have a nontrivial factor of $\rme^{K(\vp)}$ in front, and the spinor derivative of the gravitino might include a nonvanishing vector auxiliary field in \rf{Amu}.

\subsection{Advantages for cosmology}
\label{subsec:cosmoadvantages}

Supergravity actions based on \rf{actionUG} have various advantages for constructing inflationary models and for studying reheating and matter creation after inflation.  First, due to the nilpotency and orthogonality constraints, there is only one real scalar inflaton, in agreement with current cosmological data consistent with a single scalar field driving inflation.  Second, the three other scalars (complex sgoldstino and real sinflaton) which would be present for unconstrained multiplets are absent and therefore require no stabilization.  Third, the absence of the inflatino simplifies the investigation of matter creation at the end of inflation.  It is rather encouraging that these three desirable features are present in a locally supersymmetric action.

To understand this importance of these features, it is worth reflecting on the complications present in previous models.  In earlier nilpotent models with $\BS^2 = 0$ \cite{Kallosh:2014via,Dall'Agata:2014oka,Carrasco:2015uma,Carrasco:2015rva,Ferrara:2015cwa,Dall'Agata:2015zla}, the absence of the complex sgoldstino is well known.   But because $\BPhi$ was unconstrained, it was still necessary to stabilize the sinflaton field $b$ in order to ensure that the evolution of the universe is driven by a single scalar inflaton $\vp$.  For stabilizing the sinflaton, one can introduce terms associated with the bisectional curvature of the \K\ manifold \cite{Kallosh:2010xz}.  In particular, by adding terms of the form 
 \be
K \supset C(\BPhi, \BPhibar) \, \BS \BSbar (\BPhi-\BPhibar)^2,
\label{bisec} \ee
one can make the sinflaton heavy $m^2_b\sim C(\Phi, \Phibar) \, |F^s|^2$ and proportional to the moduli space bisectional curvature $R_{S\Sbar \Phi\Phibar} \sim C(\Phi, \Phibar)$.  Alternatively, terms in the \K\ potential of the form 
\be
K \supset (\BPhi- \BPhibar)^4
\label{quartic}
\ee
have been shown to stabilize $b$ \cite{Carrasco:2015uma}.  In both cases, making $b$ heavy removes it from the cosmological evolution.
 
Even if the sinflaton was constrained, however, the inflatino was still present in the dynamics, which introduced significant complications in understanding the end of inflation \cite{Kallosh:2000ve,Nilles:2001ry,Nilles:2001fg}.  To our knowledge, there is no nilpotent model in the literature with a heavy inflatino (see, however, \rf{eq:inflatinomass} below).  As a partial solution, recent inflationary models have been constructed such that $\nabla_{\phi }W=0$ at the minimum of the potential \cite{Kallosh:2014via,Dall'Agata:2014oka,Carrasco:2015uma,Carrasco:2015rva}.   In this case, there is no mixing of the gravitino with the remaining spin 1/2 field as in \rf{GI} in $\chi^s = 0$ unitary gauge, at least at the minimum.  This condition is still not quite satisfactory, though, since $\nabla_{\phi }W$ does not vanish exactly for oscillations about the minimum of the potential, still leading to gravitino-inflatino mixing.  This complicates the whole analysis of creation of matter at the end of inflation, since the equations of motion for the gravitino do not decouple from the leftover combination of the matter multiplets.  For this reason, the fact that orthogonal nilpotent multiplets have no inflatino in their spectrum is very promising.

\subsection{Gravitino dynamics}

The super-Higgs mechanism in cosmology was studied in \cite{Kallosh:2000ve}, where the supersymmetry breaking scale was shown to be equal to 
\be
 \alpha = 3 M_{\rm Pl}^2 (H^2+ m_{3/2}^2),
\ee
where $H$ is the Hubble parameter.  The equations of motion for the gravitino were derived in \cite{Kallosh:1999jj} for a model with one chiral inflaton multiplet, assuming that during inflation the inflaton is real, and therefore terms depending on the vector auxiliary field $A_{\mu}$ drop from the bosonic evolution.   Analogous equations for gravitino were given in \cite{Giudice:1999yt}.

Later, a more general form of the gravitino equations for models with any number of chiral multiplets was derived in \cite{Kallosh:2000ve}, and the case of two chiral multiplets was treated in detail.  As in section \ref{subsec:unitarygauge}, however, it was not possible to find a unitary gauge choice which would remove all mixing terms between the gravitino and the chiral fermions.  Moreover, the vector auxiliary field played a significant role in the gravitino equations of motion in an FRW background, since one had to use ``hatted''  time derivatives in the form $\hat \partial_0= \partial _0 - {i \over 2} \gamma_5 A_0^{\rm bosonic}$.

These complications are now absent for these new inflatino-less constructions.  As emphasized above, we can work in unitary gauge with $\chi^s=\chi^\phi=0$ and the vector auxiliary field is zero if we work with a \K\ potential in the form of \rf{KW}, allowing us to avoid the assumptions made in \cite{Kallosh:1999jj}.  As a result, the massive gravitino equations of motion are of the same kind as in \cite{Kallosh:1999jj,Giudice:1999yt}, since despite starting with two chiral multiplets $\BS$ and $\BPhi$, the constraints imply a unitary gauge with just a massive gravitino and no other fermions.
 
In an expanding universe, the equations of motion for the spin-1/2 and spin-3/2 components were derived in \cite{Kallosh:2000ve}. They were brought to a simpler form in \cite{Nilles:2001fg}, and in this notation, the equations for the spin-1/2 longitudinal component $\theta = \bar \gamma^i \psi_i$ and for the spin-3/2 transverse component $\vec \psi^{\,T}$ are
\begin{align}
\left [ \partial_0 + \hat B+ i \, \bar \gamma^i  k_i  \, \bar \gamma^0 \, \hat A \right]\theta(\vec{k}) &= 0, \label{gravitino-mode-spin-1/2} \\
\left [\bar \gamma^0 \partial_0 + i \, \bar \gamma^i  k_i  +  \frac{\dot a}{2} \, \bar \gamma^0 + {a \, m_{3/2}}\right]\vec \psi^{\,T}(\vec{k}) &= 0, \label{gravitino-mode-spin-3/2} 
\end{align}
where we have assumed that $m_{3/2}$ is real, $a$ is the scale factor, $\bar \gamma^\mu$ are flat space gamma matrices, dots represent derivatives with respect to physical time $\dot{f} \equiv a^{-1} \partial_0 f$, $H \equiv \dot a / a$ is the Hubble expansion rate, and 
\be
\hat{A} = -1 - \frac{2}{3} \frac{\dot{H} + \bar \gamma^0 \dot{m}_{3/2}}{H^2 + m_{3/2}^2}, \qquad \hat{B} = - \frac{3}{2} \,
\dot a \, \hat{A} + \frac{1}{2} \, a \, m_{3/2} \,
\ga^0 \left(1 + 3 \, \hat{A} \right ).
\ee
Note that the extra fermionic terms
$\Upsilon = g_j{}^i \left( \chi_i \, \partial_0 \, \phi^j + \chi^j \, \partial_0 \, \phi_i \right)$ present in the gravitino equations with two or more chiral multiplets in \cite{Kallosh:2000ve,Nilles:2001fg} are absent in our inflatino-less models since no such fermions appear in unitary gauge.

We see that the transverse modes $\vec{\psi}^{\,T}$ have canonical kinetic terms, but the spatial part of the kinetic terms for the longitudinal modes $\theta$ are modified by the matrix $\hat A$.   As recently emphasized in \cite{Kahn:2015mla}, these modifications to the kinetic structure are known as the ``slow gravitino'' \cite{Lebedev:1989rz,Kratzert:2003cr,Hoyos:2012dh,Benakli:2013ava,Benakli:2014bpa} and are required to consistently embed supersymmetry in a (cosmological) fluid background.

\section{Alternative inflatino-less constructions}
\label{sec:alternative}

Given the above discussion, it is clearly desirable to have mechanisms to decouple the inflatino from the inflationary dynamics. Here, we briefly present two alternative constructions that do not rely on the $\BS \BB = 0$ constraints, leaving a more complete study to future work.

\subsection{Linear $\BS$ terms in the \K\ potential}
\label{subsec:linearS}
 
In standard sgoldstino-less constructions with $\BS^2 = 0$ but $\BPhi$ unconstrained, the operator 
\be
\label{eq:inflatinomass}
K \supset c (\BS + \BSbar) (\BPhi - \BPhibar)^2
\ee
gives rise to an inflatino mass, where $c$ is a constant.  Note that this expression is linear in $\BS$, and the resulting inflatino mass is proportional to $c \, F^s$.  Since this term maintains the shift symmetry on $\vp$, it does not introduce any additional terms in the inflaton potential, though it does modify the dynamics of the sinflaton.  To our knowledge, this term does not appear in the literature on supergravity and inflation, though it is reminiscent of a Giudice-Masiero-like $\mu$ term \cite{Giudice:1988yz}.  It leads to a nondiagonal kinetic term $K_{S\Phibar}$ which is present during the cosmological evolution at the stage when the scalar $b(x)$ still evolves.  Note that \rf{eq:inflatinomass} does not decouple $F^\phi$ from the scalar potential, so unlike the discussion in section \ref{sec:sugrapotential}, the standard supergravity scalar potential applies to this kind of inflatino-less construction.

 \subsection{Relaxed constraints: $\BS^2=0$, $\BS \BPhibar$ is chiral}
 
 An alternative mechanism for decoupling fermions from chiral multiplets was studied in \cite{Komargodski:2009rz}, based on the constraints
 \be
 \label{eq:relaxed}
 \BS^2=0, \qquad  \Dbar_{\dot \alpha} (\BS  \BPhibar) =0,
 \ee
 such that $\BS  \BPhibar$ is a (composite) chiral multiplet.   At the level of global supersymmetry, it was shown that the first component of the inflaton multiplet $\Phi= \vp +ib$ is left unconstrained, but the expressions in \rf{inflatino} and \rf{aux} still hold, relating the inflatino and auxiliary field to $\Phi$ and the goldstino $\chi^s$.  In particular in unitary gauge with $\chi^s=0$, $\chi^\phi$ and $F^\phi$ both vanish.  Because $\Phi$ is a general complex field, these constraints do not lead to a nilpotency condition of 3rd degree ($\BB^3 \not=0$), making it possible to write down more general \K\ potentials than in the $\BS \BB = 0$ case.  We verify the above features for local supersymmetry in $\chi^s=0$ gauge in appendix \ref{app:relaxed}.
 
If we impose a shift symmetry on the \K\ potential broken only via the superpotential, these models start with
 \be
K( \BS, \BSbar;  \BPhi , \BPhibar )=  \BS \BSbar   + k(\BPhi - \BPhibar)  + c \, \BS \BSbar (\BPhi - \BPhibar)^2 + \ldots ,  \qquad W= f(\BPhi) \, \BS + g(\BPhi),
\label{KWrelaxed}
\ee  
where $k$ is a function and $c$ is a constant.  Because $F^\phi$ is constrained to be fermionic, one should again take out $\nabla_\phi W$ terms from the scalar potential as discussed in section \ref{sec:sugrapotential}.  Because the inflatino is no longer in the spectrum in unitary gauge, one need not impose $\nabla_\phi W=0$ at the minimum to avoid inflatino-gravitino mixing.  The sinflation $b$ is still in the spectrum, however, so one has to verify that it is properly stabilized.  

Models based on \rf{KWrelaxed} are interesting for studies of the initial conditions for inflation.  For example in \cite{Carrasco:2015rva}, it was found that the sinflaton evolved to the bottom of the valley at $b=0$ during cosmological evolution.  Using the relaxed constraints in \rf{eq:relaxed}, such models would effectively achieve $\BB^3=0$ dynamically, and the absence of the inflatino would still simplify the analysis of matter creation during oscillations near the minimum of the potential.

\section{Discussion}
\label{sec:discussion}

In this paper, we used orthogonal nilpotent superfields to consistently embed single-field inflation in nonlinearly realized local supersymmetry, 
thereby achieving inflatino-less models of early universe cosmology.  Perhaps a more apt description is that we achieved inflatino-sinflaton-and-sgoldstino-less cosmology, since the constraints $\BS^2=\BS \BB=0$ not only decouple the inflatino, but project out three of the four real scalars naively present in two chiral multiplets $\BS$ and $\BPhi$.   This leaves one real scalar in the spectrum with an approximate shift symmetry, yielding a locally supersymmetric action compatible with cosmological data which favors single-field slow-roll inflationary scenarios. 

Part of the reason we have emphasized the inflatino-less aspect of these constructions\footnote{(beyond the fact that ``inflatino-sinflaton-and-sgoldstino-less'' does not exactly roll off the tongue)} is that the absence of a sinflaton and sgoldstino is a feature already present in existing models.  Even for unconstrained chiral multiplets in linearly realized supersymmetry, it is known that the sectional curvature $R_{S\bar S S\bar S}$ makes the sgoldstino heavy and the bisectional curvature $R_{\Phi \bar \Phi S\bar S}$ makes the sinflaton heavy (see eqs.~(12), (13), (31) of  \cite{Kallosh:2010xz}).  There are also models based on vector multiplets in the Higgs phase \cite{Ferrara:2013rsa}, where half of a complex scalar is eaten by a massive gauge field, decoupling the sinflaton and leaving only a real inflaton at low energies.

To our knowledge, though, there is no previous supergravity construction where the inflatino does not participate in cosmological dynamics.  The fact that the inflatino is no longer an independent field---and is entirely absent in unitary gauge---is a real bonus for orthogonal nilpotent models, since it eliminates the inflatino-gravitino problem for studies of reheating after inflation \cite{Kallosh:2000ve,Nilles:2001ry,Nilles:2001fg}.  Another new feature in inflatino-less models using the orthogonal $\BS \BB = 0$ or relaxed $\Dbar_{\dot \alpha} (\BS  \BPhibar) =0$ constraints is the modified scalar potential where the contribution from $\nabla_\Phi W$ is absent.  This unique feature turns out to be very useful for building new inflatino-less cosmologies \cite{Carrasco:2015iij}. 

Before concluding, we would like add a comment here on a consistency of this scenario with gravitino scattering unitarity.\footnote{Four-fermion interactions in models with nonlinearly realized supersymmetry were studied in \cite{Brignole:1997pe}.}  In the EFT language, the constraint equations $\BS^2=\BS \BB=0$ should be regarded as ``infinitely relevant'' operators, since they impose constraints at every dynamical scale.  Because of this, one has to be careful when performing naive power counting, since the effective cutoff of these constructions is not the Planck scale, but rather the scale at which scattering amplitudes violate unitarity, in particular longitudinal gravitino scattering.  That said, the effective cutoff of the theory is generically at  $\Lambda^4 \simeq (H^2+ m_{3/2}^2) M_{\rm Pl}^2$, \cite{Kallosh:2000ve,Dall'Agata:2014oka,Kahn:2015mla}, which in most cases is safely above any of the scales considered during inflationary evolution.  We reserve a detailed study of this point to future work, where we study under which circumstances the reheating epoch can be described using constrained multiplets.  We will also study how to couple these models to matter fields and whether this might give new insights into the possible mechanisms for reheating.

It is tempting to notice that, taken at face value, these models look almost too good to be true:  a single inflaton as ordered by observations from the sky, yet the underlying action has nonlinearly realized local supersymmetry.   Based on the results of this paper, it is now possible to construct interesting and elegant cosmological models using orthogonal nilpotent superfields interacting with supergravity \cite{Carrasco:2015iij}.  We look forward to further investigations of the new observational consequences that arise from inflatino-less inflation.  

\section*{Acknowledgments}

We are grateful to  I.~Antoniadis, E.~Bergshoeff, J.~J.~Carrasco, C.~Cheung, G.~Dall'Agata, E.~Dudas, D.~Freedman, Y.~Kahn, A.~Kehagias, A.~Linde, M.~Porrati, D.~Roberts,  A.~Sagnotti,   L.~Senatore, F.~Zwirner, A.~Van Proyen and  T. Wrase for discussions and collaborations in related projects.  The work of SF is supported in part by INFN-CSN4-GSS.  The work of RK  is supported by the SITP, and by the NSF Grant PHY-1316699.  The work of JT is supported by the U.S. Department of Energy (DOE) under cooperative research agreement DE-SC-00012567, by the DOE Early Career research program DE-SC-0006389, and by a Sloan Research Fellowship from the Alfred P. Sloan Foundation.

\appendix

\section{General \K\ potential without a shift symmetry}
\label{app:generalK}

In this appendix, we show how to reduce \rf{eq:extendedK} to \rf{eq:extendedKfinal} by using field redefinitions and \K\ transformations.  Repeating \rf{eq:extendedK} for convenience, the most general \K\ potential consistent with $\BS^2 = 0$ and $\BS \BB = 0$ is
\be
\tag{\ref{eq:extendedK}}
K(\BS,  \BSbar; \BPhi ,  \BPhibar ) = h_0 (\BA) \, \BS \BSbar  + h_1(\BA) +  h_2(\BA) \, \BB  + h_3(\BA) \, \BB^2, \qquad \BA \equiv {1\over 2}\Big ( \BPhi+ \BPhibar\Big).
\ee
By the orthogonality condition $\BS \BB = 0$, though, we can freely covert $\BPhi \leftrightarrow \BPhibar$ inside of the function $h_0$:
\be
\BS \BSbar \, h_0 (\BA) = \BS \BSbar \, h_0(\BPhi) = \BS \BSbar \sqrt{h_0(\BPhi) h_0(\BPhibar)}.
\ee
Because the holomorphic field redefinition
\be
\label{eq:holofieldredef}
\BS \to \frac{\BS}{\sqrt{h_0(\BPhi)}}
\ee
does not affect the nilpotent or orthogonality conditions, we can use this freedom to set $h_0 = 1$.  By invoking the square root in \rf{eq:holofieldredef} we are implicitly assuming that $h_0(\varphi)$ is a positive function of $\varphi$ over the relevant field space, otherwise $\BS$ would have a wrong sign kinetic term.

To further simplify the \K\ potential, we can perform a general \K\ transformation on \rf{eq:extendedK} using a complex function $j$:
\be
\delta K = j(\BPhi) + j^*(\BPhibar).
\ee
Expressing this in terms of $\BPhi = \BA + i \BB$ and $\BPhibar = \BA - i \BB$ and Taylor expanding in $\BB$, we have 
\be
\delta K = \Big( j(\BA) + j^*(\BA) \Big) + i \Big( j'(\BPhi) - j^{* \prime} (\BA)  \Big) \, \BB - \Big (  j''(\BA) + j^{* \prime \prime} (\BA) \Big) \, \BB^2.
\ee
With $j$ being a real function, we can set $h_1 = 0$.  With (the derivative of) $j$ being a pure imaginary function, we can set $h_2 = 0$.  Thus, the only physical function is $h_3$, which is a generic function of $\BA$ and, in the context of cosmology, must be chosen to satisfy slow-roll requirements.

Rewriting $h_3 \to h$, the most general \K\ potential consistent with nilpotency and orthogonality is \rf{eq:extendedKfinal}, repeated for convenience:
\be
\tag{\ref{eq:extendedKfinal}}
K(\BS,  \BSbar; \BPhi ,  \BPhibar ) = \BS \BSbar + h(\BA) \, \BB^2 .
\ee
Thus, at the 2-derivative level, any orthogonal nilpotent model for cosmology depends only on the superpotential functions $f(\phi)$ and $g(\phi)$ and the \K\ potential function $h(\phi)$.

\section{Orthogonality condition in local supersymmetry}
\label{app:orthogonalitylocal}

In this appendix, we derive the local version of the $\BS \BB = 0$ constraints.  From these, it is straightforward but tedious to derive the components of the constrained $\BPhi$ superfield.  Our main goal is to support the assertion in section \ref{subsec:simplificationunitary} that the inflatino $\chi^\phi$ and auxiliary field $F^\phi$ are both zero in unitary gauge $\chi^s = 0$.

We start with multiplets written in the notation of \cite{Stelle:1978ye}:
\begin{align}
\BS &= (S, -i \chi^s_L, -i F^s, - F^s, - i \hat{D}_\mu S,0,0),\\
\BPhi &= (\varphi + i b, - i \chi^\phi_L, -i F^\phi, - F^\phi, -i \hat{D}_\mu (\varphi + i b),0,0),\\
\BPhibar &= (\varphi - i b, i \chi^\phi_R, i \Fbar^{\phi}, - \Fbar^{\phi}, i \hat{D}_\mu (\varphi - i b),0,0),\\
\BB = \frac{1}{2i} (\BPhi - \BPhibar) & = (b, -\tfrac{1}{2}\chi^\phi, -\text{Re}(F^\phi),- \text{Im}(F^\phi),-\hat{D}_\mu \varphi,0,0),\\
\BS \BB & = (C, \zeta, H, K, v_\mu, \lambda, D).
\end{align}
The components of the composite (complex) vector multiplet are:
\begin{align}
C & = S \, b, \\
\zeta & = -\tfrac{1}{2}S \chi^\phi - i b \chi^s_L, \\
H & = - S \, \text{Re}(F^\phi) - i b F^s + \tfrac{i}{4} \chibar^\phi_L \chi^s_L,\\
K & = - S \, \text{Im}(F^\phi) - b F^s + \tfrac{1}{4} \chibar^\phi_L \chi^s_L,\\
v_\mu & = - S \hat{D}_\mu \vp - i b  \hat{D}_\mu S - \tfrac{i}{4} \chibar^\phi_R \gamma_\mu \chi^s_L, \\ 
\lambda & = \tfrac{1}{2} (\slashed{\hat{D}} S + F^s) \chi^\phi_R - \tfrac{1}{2} \big(\slashed{\hat{D}} (\vp - i b) + \Fbar^{\phi} \big) \chi^s_L ,\\
D & = - i \hat{D}_\mu S \hat{D}^\mu (\vp - i b) + i F^s \Fbar^{\phi} + \tfrac{1}{4} \chibar^\phi_R \slashed{\hat{D}} \chi^s_L + \tfrac{1}{4} \chibar^s_L \slashed{\hat{D}} \chi^\phi_R.
\end{align}
The relevant supercovariant derivatives are
\begin{align}
 \hat{D}_\mu S &= \partial_\mu S - \tfrac{i}{2} \psi_\mu \chi^s_L, \\
 \hat{D}_\mu \chi^s_L & = D_\mu \chi^s_L - (\slashed{D} S) \psi_{\mu R} - F^s \psi_{\mu L} - \tfrac{i}{2} A_\mu \chi^s_L, \\
 \hat{D}_\mu \chi^s_R & = D_\mu \chi^s_R - (\slashed{D}\Sbar) \psi_{\mu L} - \Fbar^s \psi_{\mu R} + \tfrac{i}{2} A_\mu \chi^s_R,
\end{align}
where $\psi_\mu$ is the gravitino and $A_\mu$ is the vector auxiliary field of supergravity.

The $\BS \BB = 0$ constraint is now equivalent to
\be
C= \zeta=H=K=v_\mu= \lambda= D=0.
\ee
Replacing supercovariant derivatives with ordinary ones and neglecting gravitino and supergravity auxiliary fields, these reduce to the global constraints \cite{Komargodski:2009rz,Kahn:2015mla}, whose solution is \rf{scalar}-\rf{aux}.  Solving these equations in the local case is rather involved for a generic gravitino gauge choice.

In unitary gauge for the gravitino, though, $\chi^s_{L} = 0$ and $S = 0$ imply
\be
\hat{D}_\mu S \Rightarrow 0, \qquad \hat{D}_\mu \chi^s_L \Rightarrow - F^s \psi_{\mu L}, \qquad \hat{D}_\mu \chi^s_R \Rightarrow - \Fbar^s \psi_{\mu R},
\ee
as anticipated in \rf{eq:covderivunitarygauge}.  The components of $\BS \BB$ simplify dramatically in this gauge:
\be
\BS \BB \Rightarrow (0, 0, -i b F^s, - b F^s, 0, \tfrac{1}{2} F^s \chi^\phi_R, -i F^s \Fbar^{\phi}).
\ee
By assumption $F^s \not= 0$, so setting the above expression to zero implies
\be
b = 0, \qquad \chi^\phi = 0, \qquad F^{\phi} = 0,
\ee
in agreement with \rf{!} and \rf{!!}.  The only nonzero component of $\BPhi$ is $\alpha$ and we have
\begin{align}
\BS &= (0, 0, -i F^s, -F^s, 0,0,0),\\
\BPhi & = (\varphi, 0, 0, 0, -i \partial_\mu \varphi,0,0),\\
\BPhibar & = (\varphi, 0, 0, 0, i \partial_\mu \varphi,0,0).
\end{align}
Note that $\BS \BPhibar = (0,0,-i \varphi F^s, - \varphi F^s,0,0,0)$ is a chiral multiplet.

\section{Relaxed constraints in local supersymmetry}
\label{app:relaxed}

For completeness, we also consider the less restrictive constraint that $\BS \BPhibar$ is a chiral multiplet.  The calculation follows the same logic as above.  The starting point is
\begin{align}
\BS &= (S, -i \chi^s_L, -i F^s, - F^s, - i \hat{D}_\mu S,0,0),\\
\BPhibar &= (\Phibar, i \chi^\phi_R, i \Fbar^{\phi}, - \Fbar^{\phi}, i \hat{D}_\mu \Phibar,0,0),\\
\BS \BPhibar & = (C, \zeta, H, K, v_\mu, \lambda, D).
\end{align}
The components of $\BS \BPhibar$ are
\begin{align}
C & = S \, \Phibar, \\
\zeta & = i S \chi^\phi_R - i \Phibar \chi^s_L, \\
H & = i (S \Fbar^{\phi} - \Phibar F^s),\\
K & = - (S \Fbar^{\phi} + \Phibar F^s),\\
v_\mu & = i S \hat{D}_\mu \Phibar - i \Phibar  \hat{D}_\mu S - \tfrac{i}{2} \chibar^\phi_R \gamma_\mu \chi^s_L, \\ 
\lambda & = -i (\slashed{\hat{D}} S + F^s) \chi^\phi_R + i \big(\slashed{\hat{D}} \Phibar + \Fbar^{\phi} \big) \chi^s_L ,\\
D & = - 2 \hat{D}_\mu S \hat{D}^\mu \Phibar + 2 F^s \Fbar^{\phi} - \tfrac{i}{2} \chibar^\phi_R \slashed{\hat{D}} \chi^s_L - \tfrac{i}{2} \chibar^s_L \slashed{\hat{D}} \chi^\phi_R.
\end{align}

For $\BS \BPhibar$ to be chiral, we need
\be
P_R \zeta = 0, \qquad H  = i K, \qquad -i \hat{D}_\mu C = v_\mu, \qquad \lambda = D = 0.
\ee
Going to unitary gauge with $\chi^s_{L} = 0$ and $S = 0$, we have the simplification
\be
\BS \BPhibar \Rightarrow (0,0,-i \Phibar F^s,- \Phibar F^s, 0,-i F^s \chi^\phi_R, 2 F^s \Fbar^{\phi} + \tfrac{i}{2} F^s\chibar^\phi_R  \gamma^\mu \psi_{\mu L}).
\ee
Imposing the chirality constraints sets
\be
\chi^\phi = 0, \qquad F^{\phi} = 0,
\ee
but leaves the complex $\Phi$ unconstrained:
\begin{align}
\BS &= (0, 0, -i F^s, -F^s, 0,0,0),\\
\BPhi & = (\Phi, 0, 0, 0, -i \partial_\mu \Phi,0,0).
\end{align}


\end{document}